\documentclass[aps,showpacs,pra,twoside,amssymb,amsmath]{revtex4}
\usepackage{amssymb}
\usepackage{graphicx}
\usepackage{amsmath}
\usepackage{colordvi}
\usepackage{color} 
\usepackage{amsmath}
\begin{document}
\title{Exact algebraic separability criterion for two-qubit systems}
\author{Kazuo Fujikawa$^{1,2}$ and C. H. Oh$^{2,3}$\footnote{Corresponding author. Email: phyohch@nus.edu.sg.}}
\affiliation
{$^{1}$Quantum Hadron Physics Laboratory, RIKEN Nishina Center, Wako 351-0198, Japan\\
$^{2}$Institute of Advanced Studies, Nanyang Technological University, Singapore 639798, Singapore\\
$^{3}$Centre for Quantum Technologies, National University of Singapore,
Singapore 117543, Singapore}

\begin{abstract}
\noindent
A conceptually simpler proof of the separability criterion for two-qubit
systems, which is referred to  as "Hefei inequality" in literature, is presented. 
This inequality gives a necessary and sufficient separability criterion for any 
mixed two-qubit system unlike the Bell-CHSH inequality that cannot test the mixed-states such as the Werner state when regarded as a separability criterion. The original derivation of this inequality emphasized the uncertainty relation of complementary observables, but we show that the uncertainty relation does not play any role in the actual derivation and the Peres-Hodrodecki condition is solely responsible for the inequality. Our derivation, which contains technically novel aspects such as an analogy to 
the Dirac equation,
 sheds light on this inequality and on the fundamental issue to what extent the uncertainty relation can provide a test of entanglement. This separability criterion is illustrated for an exact treatment of the Werner state. 
\end{abstract}
\pacs{03.67.Mn, 03.65.Ud, 03.65.Ta}
\maketitle

\large
\section{Introduction}
The study of entanglement and nonlocality in two-qubit systems is the most fundamental as is witnessed by the
recent beautiful experimental tests of local realism~\cite{hensen, giustina, shalm, aspect}. It appears that the test of local realism in the sense of the Bell-CHSH inequality~\cite{bell, chsh} has been completed.
 Here we adopt a view that the Bell-CHSH inequality is useful as a  criterion of quantum mechanical entanglement also~\cite{acin}, and it provides a kind of the norm to which any proposed separability criterions should be compared.
We do not subscribe to a view that the role of Bell-CHSH inequality is finished once it is convincingly denied by experiments. As for the test of non-contextual hidden-variables models, which provided a basis for the Bell-CHSH inequality, those models have been mostly negated by the absence of reduction and thus causing generalized Hardy-type paradoxes; see, for example,~\cite{FYO} and references therein.

It may now be a good occasion to re-examine various separability criterions proposed in the past as tests of entanglement. The separability criterions for two-qubit systems have
been well studied, and  a comparison of various theoretical proposals has been made, for example, in Ref.~\cite{acin}.  Among various
criterions proposed in the past, the algebraic condition proposed in~\cite{yu},
which is referred to as "Hefei inequality" in~\cite{acin}, provides a necessary
and sufficient separability criterion for any mixed state and thus interesting
in view of the fact that the Bell-CHSH inequality~\cite{bell, chsh} cannot test
 general mixed states such as the Werner state~\cite{werner}.
The logical structure of the original derivation of the inequality in~\cite{yu} is, however, somewhat misleading.
 The purpose of the present paper is to show 
a conceptually simpler derivation of the inequality. To be explicit, we show
that the uncertainty relations of complementary observables do not play any role in the derivation, although the
original derivation emphasized the uncertainty relations as is indicated by its title 
``Comprehensive Test of Entanglement for Two-Level Systems via the Indeterminancy Relationship''~\cite{yu}. The discussion of the uncertainty relations introduces unnecessary complications in the analysis.  We can thus avoid the technical issue, ``It turns out that the orientation of the local testing observables plays a crucial role in our perfect detection of entanglement'' as is stated in Abstract of ~\cite{yu}.
Simply stated, the Peres-Horodecki criterion is shown to be solely responsible for their inequality, and the uncertainty relations cannot be alternative to the Peres-Horodecki criterion in the analysis of entanglement for general two-qubit systems.

\section{Derivation}

Our starting point is the fact that the general {\em pure} two-qubit states are
brought to the standard form by the Schmidt decomposition
\begin{eqnarray}
|\Phi\rangle=(u\otimes v)[s_{1}|+\rangle\otimes|-\rangle
-s_{2}|-\rangle\otimes|+\rangle]
\end{eqnarray}
with
\begin{eqnarray}
|+\rangle=\left(\begin{array}{c}
            1\\
            0
            \end{array}\right),\ \ |-\rangle=\left(\begin{array}{c}
            0\\
            1
            \end{array}\right),
\end{eqnarray}
and real numbers $s_{1}^{2}+s_{2}^{2}=1$. Namely, the states are parametrized by $s_{1}\ , s_{2}$ and two unitary matrices
$u$ and $v$.
In our analysis, we use
\begin{eqnarray}
|\Phi\rangle=(u\otimes v)[s_{1}|+\rangle\otimes|-\rangle
-s_{2}e^{i\delta}|-\rangle\otimes|+\rangle]
\end{eqnarray}
with an additional $\delta$ for technical convenience.

These states are characterized by {\em chiralities} defined by the projection operators,
\begin{eqnarray}
P_{\pm}=\frac{1}{2}(1\pm \gamma_{5})
\end{eqnarray}
with
\begin{eqnarray}
\gamma_{5}=A_{3}B_{3},
\end{eqnarray}
by defining the variables for two-qubit systems~\cite{yu},
\begin{eqnarray}
A_{k}=u\sigma_{k}u^{\dagger}\otimes 1,\ \ \ \ B_{k}=1\otimes
v\tau_{k}v^{\dagger},
\end{eqnarray}
for two sets of Pauli matrices $\{\sigma_{k}\}$ and $\{\tau_{k}\}$ using unitary
transformations $u\otimes v$ in (3).
We then have, for example, $A_{1}A_{2}=iA_{3}$.

Those operators satisfy,
\begin{eqnarray}
A_{3}(u\otimes v)|+\rangle\otimes|-\rangle=(u\otimes
v)|+\rangle\otimes|-\rangle, \ \ B_{3}(u\otimes
v)|+\rangle\otimes|-\rangle=-(u\otimes v)|+\rangle\otimes|-\rangle,
\end{eqnarray}
and thus we have
\begin{eqnarray}
P_{+}|\Phi\rangle=0, \ \ \ P_{-}|\Phi\rangle=\Phi\rangle.
\end{eqnarray}
The important property is that {\em any} pure two-qubit state is written as an
eigenstate of
the suitably chosen $P_{-}$, which simplifies the classification of pure
two-qubit states.

We start with a general pure
state
$|\Phi\rangle\langle\Phi|$ which satisfies, as above,
\begin{eqnarray}
P_{-}|\Phi\rangle\langle\Phi|P_{-}=|\Phi\rangle\langle\Phi|.
\end{eqnarray}
We then expand the state
$|\Phi\rangle\langle\Phi|$ into a complete set of $4\times 4$ Dirac matrices
\begin{eqnarray}
|\Phi\rangle\langle\Phi|=\sum_{k=1}^{16}c_{k}\Gamma_{k}
\end{eqnarray}
with $\{\Gamma_{k}\}=\{1,\ \gamma_{\mu},\gamma_{\mu}\gamma_{5},
\frac{i}{2}[\gamma_{\mu},\gamma_{\nu}], \gamma_{5}\}$,
and impose the condition $P_{-}|\Phi\rangle\langle\Phi|P_{-}=|\Phi\rangle\langle\Phi|$
for the expansion coefficients.
By this way we arrive at the expression
\begin{eqnarray}
|\Phi\rangle\langle\Phi|=\frac{1}{2}[P_{-}+a_{1}\gamma_{1}\gamma_{0}P_{-}+a_{2}\gamma_{2}\gamma_{0}P_{-}+a_{3}\gamma_{3}\gamma_{0}P_{-}]
\end{eqnarray}
with real numbers $a_{k}$ because of the hermiticity of
$|\Phi\rangle\langle\Phi|$.

Here we explicitly defined
\begin{eqnarray}
\gamma_{0}=A_{1}, \ \
\gamma_{1}=iA_{3}B_{1}, \ \
\gamma_{2}=iA_{3}B_{2}, \ \
\gamma_{3}=iA_{2},
\end{eqnarray}
which satisfy the defining relations of Dirac matrices,
\begin{eqnarray}
\{\gamma_{\mu},\gamma_{\nu}\}=2g_{\mu\nu}
\end{eqnarray}
with $g_{\mu\nu}=(1,-1,-1,-1)$.  We also have
\begin{eqnarray}\label{14}
&&\gamma_{5}=A_{3}B_{3}=-i\gamma_{0}\gamma_{1}\gamma_{2}\gamma_{3},\nonumber\\
&&\gamma_{1}\gamma_{0}=A_{2}B_{1}, \ \
\gamma_{2}\gamma_{0}=A_{2}B_{2}, \ \
\gamma_{3}\gamma_{0}=A_{3}.
\end{eqnarray}
which satisfy $\{\gamma_{5},\gamma_{\mu}\}=0$ and, for example, $[\gamma_{5},\gamma_{1}\gamma_{0}]=0$; the first relation of \eqref{14} shows that the definition of $\gamma_{5}$ in (5) is consistent with the relativistic definition.
We note that
\begin{eqnarray}
i\gamma_{k}\gamma_{0}\gamma_{5}=(\frac{1}{2})\epsilon_{klm}\gamma_{l}\gamma_{m}
\end{eqnarray}
and thus
$i\gamma_{k}\gamma_{0}\gamma_{5}P_{-}=-i\gamma_{k}\gamma_{0}P_{-}=(\frac{1}{2})\epsilon_{klm}\gamma_{l}\gamma_{m}P_{-}$;
for this reason we do not have the terms with
$(\frac{1}{2})\epsilon_{klm}\gamma_{l}\gamma_{m}P_{-}$ in the expansion in (11).

The expression (11) satisfies the condition
\begin{eqnarray}
{\rm Tr}|\Phi\rangle\langle\Phi|=1.
\end{eqnarray}
The pure state condition
\begin{eqnarray}
(|\Phi\rangle\langle\Phi|)^{2}=|\Phi\rangle\langle\Phi|
\end{eqnarray}
then implies
\begin{eqnarray}
\sum_{k}a_{k}^{2}=1.
\end{eqnarray}
Using two angles $\theta$ and $\varphi$, we thus choose
\begin{eqnarray}
|\Phi\rangle\langle\Phi|=\frac{1}{2}[P_{-}+\gamma_{3}\gamma_{0}P_{-}\cos\theta +
\gamma_{2}\gamma_{0}P_{-}\sin\theta \sin\varphi +
\gamma_{1}\gamma_{0}P_{-}\sin\theta \cos\varphi].
\end{eqnarray}

We then have the relations
\begin{eqnarray}
\langle\Phi|\rho|\Phi\rangle&=&\frac{1}{2}[\langle P_{-}\rangle_{\rho}
+\langle \gamma_{3}\gamma_{0}P_{-}\rangle_{\rho}\cos\theta\nonumber\\
&&+\langle  \gamma_{2}\gamma_{0}P_{-}\rangle_{\rho}\sin\theta \sin\varphi
+\langle \gamma_{1}\gamma_{0}P_{-}\rangle_{\rho}\sin\theta
\cos\varphi].\nonumber\\
\langle\Phi|\tau_{2}\rho^{T_{B}}\tau_{2}|\Phi\rangle&=&\frac{1}{2}[\langle
P_{+}\rangle_{\rho}
+\langle \gamma_{3}\gamma_{0}P_{+}\rangle_{\rho}\cos\theta\nonumber\\
&&-\langle  \gamma_{2}\gamma_{0}P_{-}\rangle_{\rho}\sin\theta \sin\varphi
-\langle \gamma_{1}\gamma_{0}P_{-}\rangle_{\rho}\sin\theta
\cos\varphi],
\end{eqnarray}
with
\begin{eqnarray}
\langle \gamma_{3}\gamma_{0}P_{\pm}\rangle_{\rho}\equiv{\rm
Tr}\gamma_{3}\gamma_{0}P_{\pm}\rho,
\end{eqnarray}
for example. It is interesting that $\rho$ is the operator on the left-hand side
of (20)
while $\rho$ is the state on the right-hand side. The positivity of the
"partially transposed density matrix"
\begin{eqnarray}
\rho^{T_{B}}=\sum_{ijkl}P^{ij}_{kl}|i\rangle\langle j|\otimes|l\rangle\langle k|\end{eqnarray}
for the original $\rho=\sum_{ijkl}P^{ij}_{kl}|i\rangle\langle
j|\otimes|k\rangle\langle l|$, which gives a necessary and sufficient
separability condition for $d=2\times 2$ systems~\cite{peres}, is defined here
by "partially time reversed state", $\tau_{2}\rho^{T_{B}}\tau_{2}$, for
operational simplicity~\cite{yu}. We then use
\begin{eqnarray}
\langle\Phi|\tau_{2}\rho^{T_{B}}\tau_{2}|\Phi\rangle={\rm
Tr}\rho^{T_{B}}\tau_{2}|\Phi\rangle\langle\Phi|\tau_{2}={\rm
Tr}\rho\tau_{2}\{|\Phi\rangle\langle\Phi|\}^{T_{B}}\tau_{2},
\end{eqnarray}
and
\begin{eqnarray}
\tau_{2}B^{T}_{k}\tau_{2}&=&\tau_{2}v^{\star}\tau_{2}(1\otimes
\tau_{2}\tau^{T}_{k}\tau_{2})\tau_{2}v^{T}\tau_{2}\nonumber\\
&=&-v(1\otimes \tau_{k})v^{\dagger}\nonumber\\
&=&-B_{k},
\end{eqnarray}
together with the definitions in (12) to derive the second relation in (20).

The positivity condition (Peres-Horodecki condition) of $\rho$ and
$\tau_{2}\rho^{T_{B}}\tau_{2}$ for any pure state $|\Phi\rangle\langle\Phi|$,
\begin{eqnarray}
&&\langle\Phi|\rho|\Phi\rangle\geq 0,\nonumber\\
&&\langle\Phi|\tau_{2}\rho^{T_{B}}\tau_{2}|\Phi\rangle\geq 0,
\end{eqnarray}
then implies the inequalities ({\em separability criterion})
\begin{eqnarray}
\langle P_{-}\rangle_{\rho}^{2}&\geq& \langle
\gamma_{3}\gamma_{0}P_{-}\rangle_{\rho}^{2}
+\langle  \gamma_{2}\gamma_{0}P_{-}\rangle_{\rho}^{2}
+\langle  \gamma_{1}\gamma_{0}P_{-}\rangle_{\rho}^{2},\nonumber\\
\langle P_{+}\rangle_{\rho}^{2}&\geq&
\langle \gamma_{3}\gamma_{0}P_{+}\rangle_{\rho}^{2}+\langle
\gamma_{2}\gamma_{0}P_{-}\rangle_{\rho}^{2}+\langle
\gamma_{1}\gamma_{0}P_{-}\rangle_{\rho}^{2},
\end{eqnarray}
by noting
\begin{eqnarray}
&&{\rm Tr}\{|\Phi\rangle\langle\Phi|\rho\}\\
&&=\frac{1}{2}\left[\langle P\rangle_{\rho}\nonumber
+
\sqrt{\langle(\gamma_{3}\gamma_{0}P)\rangle^{2}_{\rho}+[\langle(\gamma_{1}\gamma_{0}P)\rangle_{\rho}^{2}+\langle(\gamma_{2}\gamma_{0}P)\rangle_{\rho}^{2}]\cos^{2}(\varphi+\beta)}\cos(\theta+\alpha)\right],\nonumber
\end{eqnarray}
where we used the general formula of trigonometric functions $A\cos\varphi+B\sin\varphi=\sqrt{A^{2}+B^{2}}\cos(\varphi+\beta)$ with suitable $\beta$ in (20) to derive (27); the conditions (25) should hold for any $\varphi$ and $\theta$ with fixed
$\alpha$ and $\beta$.  For the case of the first relation in (20) with
$P=P_{-}$, one obtains the first relation of (26) by choosing $\cos(\theta+\alpha)=-1$ and $\cos^{2}(\varphi+\beta)=1$ in (27). Similarly one obtains the second relation in (26) from (20). The relations (20) are
equalities and thus those conditions (26), if imposed for {\em any} $u$ and $v$,
provide a {\em necessary and sufficient separability condition}.  We mention that the information about the explicit parametrization
of the initial state in (3) is lost after this operation of extremalizing the expression except for $P_{-}|\Psi\rangle=|\Psi\rangle$.  Note also that
$\langle P\rangle_{\rho}\geq 0$ because $P=P^{2}$. 

It is interesting that the first relation in (26) formally states that the 4-vector $\langle
\gamma_{\mu}\gamma_{0}P_{-}\rangle_{\rho}$ is a time-like vector. The 4-dimensional Dirac notation is used in the present analysis just to make the counting of all the possible terms in the expansion (19) straightforward and that the Peres-Horodecki condition is neatly expressed by the chiral projection operators $P_{\pm}$. The Peres-Horodecki condition changes the projection operator $P_{-}$ in the first relation of (26), which always holds in quantum mechanics, to $P_{+}$ in two of the terms in the second relation. If one changes all $P_{-}$ to $P_{+}$, one obtains another relation which always holds; this new relation is derived by starting with the Schmidt decomposition $(u\otimes v)[s_{1}|+\rangle|+\rangle-s_{2}e^{i\delta}|-\rangle|-\rangle]$. 
\\

The relations in (26) corespond to the separability criterions of Ref.~\cite{yu}. It should however be emphasized that we never refer to uncertainty relations of complementary local observables in our derivation of (26), unlike 
the original derivation which follows the idea ``A 3-setting Bell-type inequality enforced by the indeterminancy relation of complementary local observables is proposed as an experimental test of 2-qubit entanglement'' as is stated in Abstract of ~\cite{yu}. 
It should be mentioned that the derivation in~\cite{yu} used the Peres-Horodecki condition also in addition to uncertainty relations. Our relations (26) formally appear to be different from the relations in ~\cite{yu}, since our relations contain one more term on the right-hand side, namely, the last terms. Our relations may thus appear to be stronger than the original relations in~\cite{yu}, but the actual applications below show that they are practically equivalent. 
The first relation in (26) should always hold for any
sensible positive definite density matrix.

The second relation in (26) provides a separability criterion.
For example, for a given pure state $\rho=|\psi\rangle \langle\psi|$ with a
representation,
\begin{eqnarray}
|\psi\rangle=(u_{0}\otimes v_{0})\frac{1}{\sqrt{2}}[|+-\rangle
-|-+\rangle]\equiv(u_{0}\otimes v_{0})|\psi_{0}\rangle,
\end{eqnarray}
the first relation in (26) should always hold, while the second relation {\em
can be violated}.
By choosing arbitrary $u$ and $v$ at $u=u_{0}$ and $v=v_{0}$, and noting
$P_{+}\rho=0$, the second relation in (26) becomes
\begin{eqnarray}
0&\geq&\langle \gamma_{2}\gamma_{0}P_{-}\rangle_{\rho}^{2}+\langle
\gamma_{1}\gamma_{0}P_{-}\rangle_{\rho}^{2},
\end{eqnarray}
which generally fails, and in fact, $\langle
\gamma_{2}\gamma_{0}P_{-}\rangle_{\rho}^{2}=\langle\psi_{0}|\sigma_{2}\otimes\tau_{2}|\psi_{0}\rangle^{2}$
is related to the notion of concurrence~\cite{wootters}, and $\langle
\gamma_{2}\gamma_{0}P_{-}\rangle_{\rho}^{2}>0$. The state (28) is thus inseparable.

If one wants to analyze pure inseparable states based on the specific state
\begin{eqnarray}
|\psi_{0}^{\prime}\rangle=\frac{1}{\sqrt{2}}[|++\rangle -|--\rangle],
\end{eqnarray}
for example, one may choose $u_{0}$ and $v_{0}$, which include a rotation of the
second state in $[|++\rangle -|--\rangle]/\sqrt{2}$ by 180 degrees around the
axis 2, and then the analysis is reduced to that of (28).

For {\em any pure separable state} $\rho=|\psi\rangle \langle\psi|$ with,
 for example, $|\psi\rangle=(u_{0}\otimes v_{0})|+-\rangle$, one can confirm
that two relations in the criterion (26) are simultaneously satisfied: To see
this, one first observes that
the first relation in (26) generally holds for any positive normed states
$|\Phi\rangle$.
On the other hand,
$(i\tau_{2})\rho^{T_{B}}(-i\tau_{2})=|\psi^{\prime}\rangle\langle\psi^{\prime}|$
with $|\psi^{\prime}\rangle=(u_{0}\otimes v_{0})
|++\rangle$ defines a physical density matrix, and thus the second condition in
(26)
for any positive normed states $|\Phi\rangle$ is satisfied. This argument is
generalized to a general mixed state
\begin{eqnarray}
\rho=\sum_{k}w_{k}\rho_{k}
\end{eqnarray}
with separable pure states $\rho_{k}$. It is thus seen that the separability
conditions (26) provide a necessary condition for mixed states. Our derivation
following Peres and Horodecki shows that the relations (26) also provide a
sufficient separability condition for mixed states.\\

As for the test of the Werner state~\cite{werner} which is defined by
\begin{eqnarray}
\rho_{w}=\frac{1}{4}(1-\beta){\bf 1} +\beta|\psi_{s}\rangle\langle\psi_{s}|
\end{eqnarray}
with the singlet state
\begin{eqnarray}
|\psi_{s}\rangle=(1/\sqrt{2})[|+\rangle|-\rangle-|-\rangle|+\rangle],
\end{eqnarray}
one can use the second relation in (26).  Namely,
\begin{eqnarray}
\langle P_{+}\rangle_{\rho_{w}}\geq|\langle\gamma_{2}\gamma_{0}P_{-}\rangle_{\rho_{w}}|,
\end{eqnarray}
since the two other terms on the right-hand side of the second relation in (26) are positive semi-definite; in fact, they vanish
for the Werner state if one sets $u=v=1$ in (6) by noting $\gamma_{3}\gamma_{0}P_{+}=(A_{3}+B_{3})/2$ and $\gamma_{1}\gamma_{0}P_{-}=(A_{2}B_{1}-A_{1}B_{2})/2$,
and thus our analysis of the bound on $\beta$ below is exact.
The relation (34)  is rewritten as
\begin{eqnarray}
1\geq -\langle \gamma_{5}\rangle_{\rho_{w}}+|\langle\gamma_{2}\gamma_{0}(1-\gamma_{5})\rangle_{\rho_{w}}|,
\end{eqnarray}
or more explicitly
\begin{eqnarray}
1\geq -\langle A_{3}B_{3}\rangle_{\rho_{w}}+|\langle A_{1}B_{1}+A_{2}B_{2}\rangle_{\rho_{w}}|,
\end{eqnarray}
which gives
\begin{eqnarray}
1\geq \beta +2\beta,
\end{eqnarray}
by setting $u=v=1$ in (6); $3\beta$ is also the upper bound to the operator norm of the right-hand side of (36) for the Werner state (32).
We thus conclude that the separability condition of the Werner state is  equivalent to
\begin{eqnarray}
\beta \leq \frac{1}{3},
\end{eqnarray}
which agrees with the result of a more explicit analysis of $\rho_{w}$~\cite{werner}.  This in particular implies that  $\beta > \frac{1}{3}$ stands for an {\em inseparable state}, which is established by (34) without using $\langle \gamma_{3}\gamma_{0}P_{+}\rangle_{\rho_{w}}^{2}=\langle
\gamma_{1}\gamma_{0}P_{-}\rangle_{\rho_{w}}^{2}=0$; what our precise analysis establishes is that $\beta = (1-\epsilon)/3$ with
$\epsilon \geq 0$ gives a separable state. The commonly used value $\beta=1/2$ defines an  inseparable Werner state although it satisfies
the CHSH inequality~\cite{werner}.  See also Ref.~\cite{fujikawa2} for a discussion of the specific properties of two-photon
states, which requires a careful analysis.

\section{Discussion}
Recent experiments~\cite{hensen, giustina, shalm} show that the very sophisticated tests of local realism in the sense of Bell-CHSH inequality are now available. The test of entanglement however covers not only the local realism but also the quantum mechanical separability, and thus  any inequality which gives a precise criterion for any mixed state is interesting.
We re-analyzed one of such inequalities proposed in~\cite{yu},
and we have shown that the separability conditions in (26) are derived without referring to uncertainty relations for complementary variables which were emphasized in the original derivation~\cite{yu}. This greatly
simplified a logical structure of the
derivation. (As for the possible implications  of  uncertainty
relations for a sufficient number of operators,
see, for example,~\cite{yu2}). To be precise, the original derivation in~\cite{yu} emphasized uncertainty relations of complementary observables as essential ingredients but at the same time used the Peres-Horodecki condition in the actual analysis, while the present derivation clarified the basic conceptual aspect by showing that only the Peres-Horodecki condition is needed. This shows that the uncertainty relations cannot be alternative to the Peres-Horodecki condition in the analysis of entanglemednt for general two-qubit systems.

The practical applications of the separability condition in~\cite{yu} have been discussed in~\cite{acin}, and its advantages and disadvantages have been  mentioned.  A much simplified logical structure of the derivation of inequalities (26), namely, a clear statement of the essential ingredient to derive (26)  
will make the inequalities useful in practical applications. As a concrete appliction, we mention the recent analysis of entanglement by means of the angular-averaged two-point correlations in two-qubit systems~\cite{fujikawa2}, which can test the Werner state without recourse to the state reconstruction,
and the criterion (26) provided very useful information in this analysis.  We expect more similar applications in the future.  The ``Hefei inequality'' stands for
a rare algebraic criterion that is valid for any mixed state which cannot be tested by the Bell-CHSH inequality
 in general~\cite{bell, chsh},  as was illustrated by an exact treatment of the Werner state,  and it  is our opinion that the
 inequalities in (26) deserve more attention.
\\

In comparison with the criterions of separability of two-qubit systems, we here briefly mention a corresponding test of the separability of systems with two continuous degrees of
freedom~\cite{simon, duan, fujikawa}. In the problem of two-party continuum case with two-dimensional
continuous phase space freedom  $(p,q)$ in each party
as defined by Simon~\cite{simon} and Duan et al.~\cite{duan}, it is possible to
re-formulate the problem such that~\cite{fujikawa}; \\
(i) the uncertainty relation leads to a necessary condition for separable
two-party systems,\\
(ii) the derived condition is sufficient to prove the separability of two-party
Gaussian systems.

Namely, the uncertainty relation {\em without} referring to the Peres-Horodecki
criterion~\cite{peres} provides a
necessary and sufficient separability condition for two-party Gaussian systems.
Simon~\cite{simon} formally used the Peres-Horodecki criterion, but an even
stronger condition is obtained in the case of continuum from the analysis of Kennard's type
uncertainty relations~\cite{fujikawa} without referring to the Peres-Horodecki
criterion.
This derivation of the necessary and sufficient condition without referring to the Peres-Horodecki
criterion is possible because we
concentrate on the case of specific two-party Gaussian systems in~\cite{simon, duan, fujikawa}.  In other words, the genuine use of the Peres-Horodecki criterion for the two-party systems with continuous freedom may give rise to useful separability criterions for more general systems than Gaussian systems.

\section*{Acknowledgments}
We thank Sixia Yu for numerous clarifying comments on the Hefei
inequality.
One of the authors (K.F.) thanks K-K. Phua for the hospitality at IAS, Nangyang
Technological University.
This work is partially supported by JSPS KAKENHI (Grant No. 25400415) and the
National Research Foundation and Ministry of Education, Singapore (Grant No.
WBS: R-710-000-008-271).

\end{document}